\documentclass[aps,prl,numerical,linenumbers,superscriptaddress,showpacs,floatfix,reprint]{revtex4-1}
\usepackage[dvipdfm]{graphicx}
\usepackage{amsmath,amssymb,amsfonts}
\usepackage{color}

\newcommand{\be}{\begin{eqnarray}}
\newcommand{\ee}{\end{eqnarray}}

\def\v2{\mbox{$v_2$}}

\newcommand{\mean}[1]{\left\langle #1 \right\rangle}

\begin{document}
\title{Scaling patterns for the suppression of charged hadron yields in Pb+Pb collisions \\at $\sqrt{s_{NN}}=2.76$~TeV:
Constraints on transport coefficients}
%
\author{ Roy~A.~Lacey} 
\author{ N.~N.~Ajitanand} 
\author{ J.~M.~Alexander}
\author{ J.~Jia}
\author{A.~Taranenko}
%
\affiliation{Department of Chemistry, 
Stony Brook University, \\
Stony Brook, NY, 11794-3400, USA}


\date{\today}
\begin{abstract}

	Suppression measurements for charged hadrons are used to investigate the 
path length ($L$) and transverse momentum ($p_T$) dependent jet quenching patterns of 
the hot and dense QCD medium produced in Pb+Pb collisions at $\sqrt{s_{NN}}=2.76$\,TeV at 
the LHC. The observed scaling patterns, which are similar to those observed for Au+Au 
collisions at $\sqrt{s_{NN}}=0.20$\,TeV at RHIC, show the  
trends predicted for jet-medium interactions dominated by radiative energy loss. 
They also allow a simple estimate of the transport coefficient $\hat{q}$, which 
suggests that the medium produced in LHC collisions is somewhat less opaque than that 
produced at RHIC, if the same parton-medium coupling strength is assumed. 
The higher temperature produced in LHC collisions could reduce the  parton-medium 
coupling strength to give identical values for $\hat{q}$ in LHC and RHIC collisions.

\end{abstract}
\maketitle


During the early stage of a relativistic heavy ion collision, 
quarks and gluons are often scattered to large transverse momentum
$p_T$. These scattered partons can interact and lose energy as they traverse 
the short-lived quark-gluon plasma (QGP), also produced in the collision \cite{Gyulassy:1993hr}.
The scattered partons which subsequently emerge, fragment into topologically aligned high-$p_T$ 
hadrons or jets; their suppressed yields encode the degree of parton energy loss. 
This essential consequence of parton energy loss has been characterized at the Relativistic 
Heavy Ion Collider (RHIC), via the observation that high-$p_T$ hadron yields are 
suppressed in central and mid-central AA collisions, in comparison to the binary-scaled 
yields from p+p collisions \cite{Adcox:2001jp,Adler:2002xw}.
The magnitude and trend of this suppression -- termed ``jet quenching'' -- has 
been a key ingredient in recent attempts to estimate the transport properties of 
the QGP \cite{Qin:2007rn,Bass:2008rv,Lacey:2009ps,Sharma:2009hn,CasalderreySolana:2010eh,
Qin:2010mn,Renk:2011gj,Chen:2011vt,Majumder:2011uk,Zakharov:2011ws,Betz:2012qq}. 

	The Large Hadron Collider (LHC) has now extended the available
c.m. energy range for AA collisions by more than a factor of ten, enabling 
investigations of the energy loss of much more energetic partons in the QGP medium produced 
at a higher temperature \cite{Aamodt:2010jd,Appelshauser:2011ds,CMS:2012nt, Milov:2011jk}. 
Relative to Au+Au collisions at RHIC ($\sqrt{s_{NN}} = 0.2$ TeV), the measured 
multiplicity for Pb+Pb collisions at the LHC ($\sqrt{s_{NN}} = 2.76$ TeV) suggests an 
approximately 30\% increase in the temperature of the QGP medium.	
This increase in temperature could result in a lowering of the strong interaction coupling 
strength, as well as a change in the stopping power of the medium. Either could 
have a significant influence on the magnitude of parton energy loss, which in turn,  
influences the magnitude and trend of jet quenching. 
Thus, an important open question is the extent to which jet quenching measurements
differ at RHIC and the LHC, and whether a characterizable difference gives an 
indication for the expected change in the properties of the medium created 
in LHC collisions.
 

%
\begin{figure*}[t]
\includegraphics[width=1.0\linewidth]{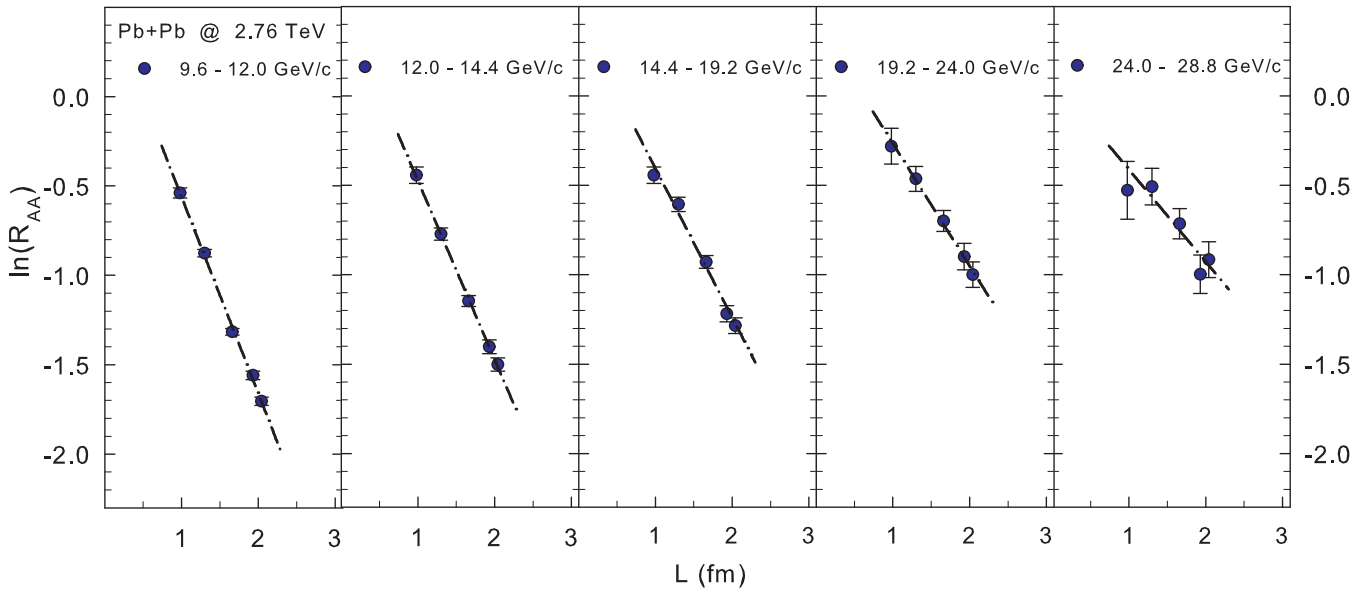}
\caption{ (Color online) $\ln\left[R_{\rm AA}(p_T,L)\right]$ vs. $L$ for several 
$p_T$ selections as indicated. Error bars are statistical only. 
The dot-dashed curve in each panel is a linear fit to the data (see text).
} 
\label{Fig1}
\end{figure*}

	An experimental probe commonly used to quantify jet quenching in AA
collisions is the nuclear modification factor ($R_{\rm AA}$);
\[
   R_{\rm AA}(p_T) = \frac{1/{N_{\rm evt}} dN/dydp_{\rm T}}{\mean{T_{\rm AA}} d\sigma_{pp}/dydp_{\rm T}}, 
\]
where $\sigma_{pp}$ is the particle production cross section in p+p collisions 
and $\mean{T_{\rm AA}}$ is the nuclear thickness function
averaged over the impact parameter ($\mathbf{b}$) range associated with a given 
centrality selection
\[
\langle T_{AA}\rangle\equiv
\frac {\int T_{AA}(\mathbf{b})\, d\mathbf{b} }{\int (1- e^{-\sigma_{pp}^{inel}\, T_{AA}(\mathbf{b})})\, d\mathbf{b}}.
\]
The average number of nucleon-nucleon collisions, 
$\langle N_{coll}\rangle=\sigma_{pp}^{inel} \langle T_{AA}\rangle$,
is usually obtained via a Monte-Carlo Glauber-based 
model calculation \cite{Miller:2007ri,Alver:2006wh}.

The $R_{\rm AA}$ measurements for neutral pions ($\pi^0$) produced in RHIC collisions
($\sqrt{s_{NN}} = 0.2$ TeV), show a characteristic dependence on both $p_T$ and 
the estimated path length $L$ of the medium. 
More specifically, $\ln\left[R_{\rm AA}(p_T,L)\right]$ has been observed to scale 
as $L$ and $1/\sqrt(p_T)$ respectively ({\em i.e.} $\ln\left[R_{\rm AA}(p_T,L)\right]$ shows a 
linear dependence on $1/\sqrt(p_T)$ for fixed values of $L$, and a linear dependence on $L$ 
for fixed values of $p_T$.) \cite{Lacey:2009ps}. These scaling patterns reflect the predicted 
quenching of the transverse momentum spectrum for jets produced from scattered light partons 
which loose energy via medium induced gluon radiation \cite{Dokshitzer:2001zm};
\begin{eqnarray}
R_{\rm AA}(p_T,L) \simeq \exp \left[- {2 \alpha_s C_F \over \sqrt{\pi}}\ 
L\,\sqrt{\hat{q}\frac{{\cal{L}}}{p_T}}\, \right] \nonumber \\
{\cal{L}} \equiv \frac{d}{d\ln p_T} 
\ln \left[ {d \sigma_{pp} \over d p_{T}^2}( p_{T})\right], 
\label{eq:DK1}	
\end{eqnarray}
where $\alpha_s$ is the strong interaction coupling strength, $C_F$ is the color factor 
and $\hat{q}$ is the transport coefficient which characterizes 
the squared average transverse momentum exchange [per unit path length] between the medium 
and the parton. 

The excellent agreement between the scaling patterns observed for the RHIC $R_{\rm AA}$ data 
and those predicted by Eq.~\ref{eq:DK1}, has been interpreted as an indication that 
medium induced gluon radiation dominates the underlying mechanism for jet quenching 
in RHIC collisions \cite{Lacey:2009ps}. The slopes of these scaling curves, which encode the magnitude 
of $\alpha_s$ and $\hat{q}$ (cf. Eq.~\ref{eq:DK1}), have also been used to extract 
a simple estimate of $\hat{q}$ for the medium produced in these collisions \cite{Lacey:2009ps} .

	A fundamental change in the mechanism for jet quenching is not expected as 
the beam energy is raised from $\sqrt{s_{NN}} = 0.2$ (RHIC) 
to $\sqrt{s_{NN}} = 2.76$ TeV (LHC). 
Thus, the scaling patterns observed for RHIC $R_{\rm AA}$ data, might
also be expected at the higher beam collision energy. We therefore search for these 
scaling patterns in LHC $R_{\rm AA}$ data, with an eye towards a possible 
difference in the slopes of the scaling curves for the two beam energies.
Such slope differences could signal a change in the properties of the medium created 
in LHC collisions.

	The $R_{\rm AA}$ measurements employed in our search were  
recently reported for charged hadrons by the CMS collaboration \cite{CMS:2012nt}. 
These data and their associated errors, are shown as a function of $p_T$ 
for several centrality selections in Fig.~4 of Ref. \cite{CMS:2012nt}. 
They indicate that suppression is modest in peripheral collisions, but is 
increasingly pronounced in more-central collisions, as might be expected from  
the longer path lengths associated with central and 
mid-central collisions (cf. Eq.~\ref{eq:DK1}). They also indicate that 
$R_{\rm AA}$ reaches a centrality dependent minimum value for 
$p_T \approx 6-7$ GeV/c, but shows a clear increase with $p_T$ up to at least 40 GeV/c. 
These features provide the substance for the scaling search discussed below.

To facilitate comparisons to our earlier scaling analysis of RHIC $\pi^0$ data,  
we apply the cut $p_T \agt 10$~GeV/c for our scaling search. Here, it is noteworthy 
that RHIC measurements indicate that the $R_{\rm AA}$ values for 
neutral pions and charged hadrons converge for 
$p_T \agt 9$ GeV/c, \cite{Adare:2008qa,Adams:2003kv}. We also use the transverse size of the 
system $\bar{R}$ as a simple estimate for the path length $L$, as was done in 
our earlier analysis. A Monte-Carlo Glauber-based model 
calculation \cite{Miller:2007ri,Alver:2006wh} 
was used to evaluate the values for $\bar{R}$ in Pb+Pb collisions as follows.
For each centrality selection, the number of participant nucleons 
$N_{\rm part}$, was first estimated. Subsequently, $\bar{R}$ was determined from the 
distribution of these nucleons in the transverse ($x,y$) plane as:
%
$
{1}/{\bar{R}}~=~\sqrt{\left(\frac{1}{\sigma_x^2}+\frac{1}{\sigma_y^2}\right)},
%
%
%
$
%
where $\sigma_x$ and $\sigma_y$ are the respective root-mean-square widths of
the density distributions.  
For these calculations, the initial entropy profile in the transverse
plane was assumed to be proportional to a linear combination
of the number density of participants and binary collisions \cite{Hirano:2009ah,Lacey:2009xx}.
The latter assures that the entropy density weighting used, is constrained by the Pb+Pb hadron 
multiplicity measurements \cite{Chatrchyan:2011pb}. 
Averaging for each centrality, was performed over the configurations generated in the 
simulated collisions.
\begin{figure*}[t]
\includegraphics[width=0.95\linewidth]{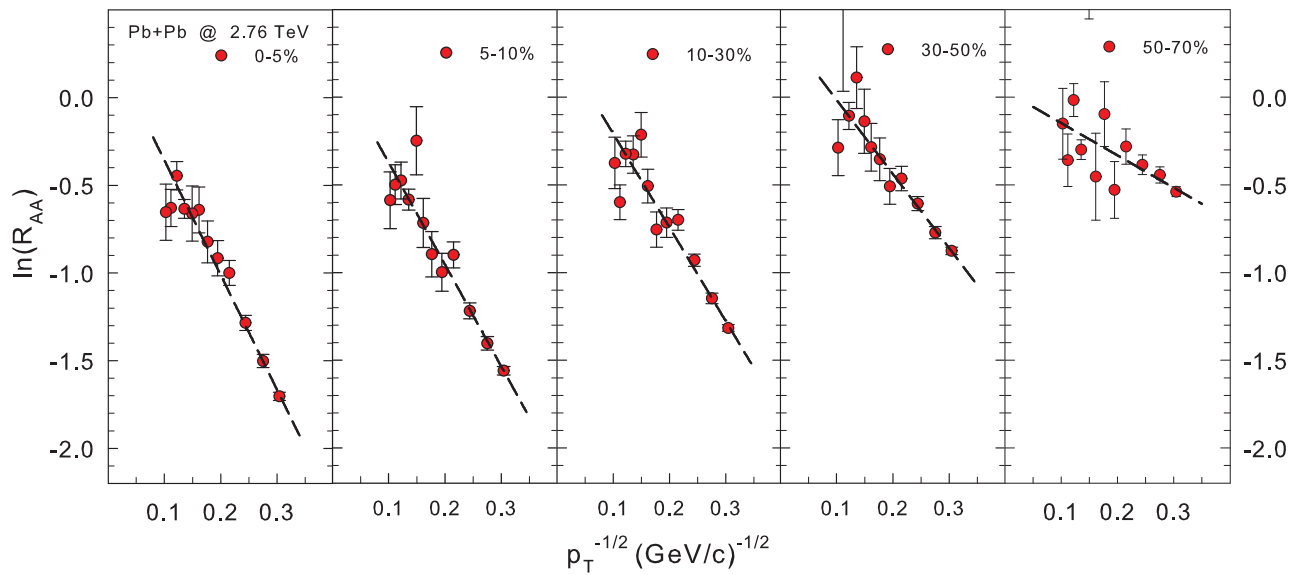}
\caption{(Color online) $\ln\left[R_{\rm AA}(p_T,L)\right]$ vs. $1/\sqrt(p_T)$ ($p_T \agt 10$~GeV/c) 
for several centrality selections as indicated. Error bars are statistical only. 
The dashed curve in each panel is a fit to the data (see text).
} 
\label{Fig2}
\end{figure*}

	The main results from our scaling search are presented in Figs. \ref{Fig1} and \ref{Fig2}. 
The plots of $\ln\left[R_{\rm AA}(p_T,L)\right]$ vs. $L$ are shown for different $p_T$
selections [as indicated] in the five panels shown in Fig.~\ref{Fig1}. Within errors, all of these plots 
show the linear increase with $L$, predicted in Eq.~\ref{eq:DK1}. This linear dependence is exemplified 
by the dashed-dot curves which represent linear fits to the data. 
The fits indicate an intercept $L \approx 0.5 \pm 0.1$~fm (for $\ln\left[R_{\rm AA}(p_T,L)\right] = 0$), 
which is similar to the value $L \approx 0.6 \pm 0.1$~fm observed in RHIC collisions \cite{Lacey:2009ps}. 
This suggests a similar minimum path length requirement for the initiation of jet quenching 
in RHIC and LHC collisions. 
Note that this requirement is akin to the plasma formation or cooling times proposed in 
Refs.~\cite{Pantuev:2005jt,Liao:2005hj}. The slopes $S_L$ of the curves in Fig. \ref{Fig1} also 
show a decrease with increasing $p_T$ selection, indicating the expected mild decrease in jet 
quenching as $p_T$ is increased (cf. Eq.~\ref{eq:DK1}).	
	
The complimentary plots of $\ln\left[R_{\rm AA}(p_T,L)\right]$ vs. $1/\sqrt(p_T)$ are shown in 
Fig. \ref{Fig2} for five separate centrality selections as indicated. In likeness to Fig. \ref{Fig1},
all of these plots show the predicted linear decrease with $1/\sqrt(p_T)$ (cf. Eq.~\ref{eq:DK1}); here, 
the dashed curves represent linear fits to the data. The extrapolated intercepts 
($\ln\left[R_{\rm AA}(p_T,L)\right] = 0$) 
of these curves indicate that the linear decrease of jet quenching persists up to the relatively high 
value $p_T \approx 800$~GeV/c in central collisions.  The corresponding slopes $S_{p_T}$ of 
these curves  decrease as collisions become more peripheral, indicating the 
effects of the shorter path lengths that partons traverse as 
collision become more peripheral.

\begin{figure}[t]
\includegraphics[width=1.0\linewidth]{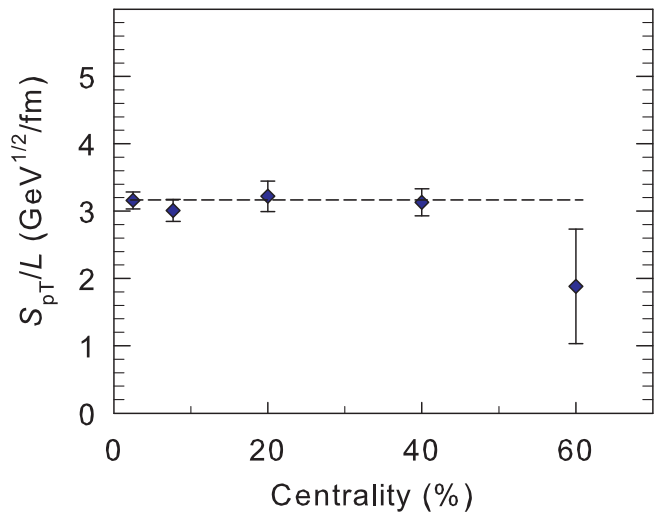}
\caption{(Color online) Centrality dependence of $S_{p_T}/L$, see text. 
The slopes $S_{p_T}$ are obtained from the the linear fits shown in Fig. \ref{Fig2}.
The dashed curve is drawn to guide the eye.
} 
\label{Fig4}
\end{figure}

	For a given medium (fixed $\hat{q}$) Eq.~\ref{eq:DK1} suggests that the ratio 
${S_{p_T}}/{L}$ should be independent of the collision centrality, and the 
product $S_L \sqrt(p_T)$ should be independent of $p_T$.  
The flat centrality dependence shown for ${S_{p_T}}/{L}$, in Fig. \ref{Fig4},
validates this prediction. Within errors, the product $S_L \sqrt(p_T)$ shows 
a similarly flat $p_T$ dependence for $p_T$ values which allow reliable extraction 
of $S_L$. The magnitude of the values for $S_L \sqrt(p_T)$ are also similar to 
those for ${S_{p_T}}/{L}$, as expected from Eq.~\ref{eq:DK1}. These observations 
all suggest the validity of Eq.~\ref{eq:DK1}. Therefore, we use the average value 
of these ratios and products to estimate $\hat{q}_{LHC}$ and compare it to 
the $\hat{q}_{RHIC}$ value previously extracted with the same technique. 

	One can compare the values for $\hat{q}$ at $\sqrt{s_{NN}} = 0.2$ and 2.76~TeV as follows: 
First, we average the values of of ${S_{p_T}}/{L}$ and $S_L \sqrt(p_T)$ to obtain the 
value $3.3 \pm 0.15$ GeV$^{1/2}$/fm. Then using Eq.~\ref{eq:DK1} with values of 
$\alpha_s = 0.3$ \cite{Bass:2008rv}, $C_F = 9/4$ \cite{cfactor,Dokshitzer:2001zm} 
and ${\cal{L}} = n = 6.7$ \cite{ppPower}, one obtains 
$\hat{q}_{LHC} \approx 0.56 \pm 0.05$~GeV$^2$/fm. 
This estimate of $\hat{q}_{LHC}$, which can be interpreted as a space-time average, 
is approximately 25\% smaller than our earlier estimate 
of $\hat{q}_{RHIC} \approx 0.75 \pm 0.05$~GeV$^2$/fm, 
evaluated with the same values for $C_F$ and $\alpha_s$ \cite{Lacey:2009ps}. 
Thus, the QCD medium produced in LHC collisions, seems to be somewhat less opaque 
to partons than the medium produced at RHIC, if a fixed value of the parton-medium coupling strength 
is assumed \cite{Betz:2012qq}. However, since the exponent of Eq.~\ref{eq:DK1} varies as $\alpha_s \sqrt{\hat{q}}$,
it is the product $\alpha_s^2{\hat{q}}$ that is $\sim 25$\% smaller for $\sqrt{s_{NN}}=2.76$~TeV.
Therefore, a small ($\sim 12.5\%$) thermal suppression of $\alpha_s$ would lead 
to identical magnitudes for $\hat{q}_{LHC}$ and $\hat{q}_{RHIC}$.
We conclude that the value of the transport coefficient $\hat{q}$, is very similar for 
the hot and dense medium created in RHIC and LHC collisions, although a possible 
change in $\alpha_s$ has not been independently established. The close similarity between 
$R_{\rm AA}$ measurements at RHIC and the LHC is a good indicator 
that $\hat{q}_{LHC}$ and $\hat{q}_{RHIC}$ are quite comparable.

	In summary, we have performed validation tests of the scaling properties 
of jet suppression in Pb+Pb collisions at $\sqrt{s_{NN}} = 2.76$ TeV. 
These tests confirm the $1/\sqrt(p_T)$ dependence, as well as the linear dependence on 
path length predicted by Dokshitzer and Kharzeev for jet suppression dominated 
by the mechanism of medium-induced gluon radiation in a hot and dense QGP. 
The quenching patterns indicate a minimum path length requirement 
for the initiation of charged hadron suppression, but suggest that jet quenching 
extends up to a relatively high value of $p_T$ in central collisions. 
For a fixed value of the coupling strength $\alpha_s$, 
the QCD medium produced in LHC collisions appears to be less opaque 
to partons than at RHIC. However, a small ($\sim 12.5\%$) thermal suppression of $\alpha_s$ 
would lead to the same magnitude for $\hat{q}_{LHC}$ and $\hat{q}_{RHIC}$. Such a suppression
might result from the approximately 30\% growth in the temperature of the QGP medium
produced in LHC collisions. The extracted values for $\hat{q}_{{LHC}}$ and $\hat{q}_{RHIC}$
are comparable to the recent estimates of $\sim 1 - 2$~GeV$^2$/fm
obtained from fits to hadron suppression data within the framework of 
the higher twist (HT) expansion \cite{Zhang:2007ja,Majumder:2011uk} and the Gyulassy-Levai-Vitev (GLV) 
scheme \cite{Gyulassy:2000fs,Betz:2012qq}.
However, they are much smaller than the value extracted 
via the approach of Arnold, Moore and Yaffe (AMY) \cite{Bass:2008rv,Qin:2007zzf} and that of 
Armesto Salgado and Wiedemann (ASW) \cite{Bass:2008rv,CasalderreySolana:2010eh}.
These results should provide important model constraints in ongoing attempts to use jet-quenching 
as a quantitative tomographic probe of the QGP.

\section*{Acknowledgments}
We thank Yen-Jie Lee for crucial communications about CMS data.
This work was supported by the US DOE under contract DE-FG02-87ER40331.A008.
 

\bibliography{LHC_qhatRefs}

\begin{thebibliography}{33}
\bibitem [{cfa()}]{cfactor}%
  \BibitemOpen
  \href@noop {} {}\bibinfo {note} {Here, it is assumed that charged hadrons are
  dominantly produced via the fragmentation of gluons.}\BibitemShut {Stop}%
\end{thebibliography}
\end{document}